\documentclass[twocolumn,letter]{jpsj3}
\usepackage{amsmath}
\usepackage{amssymb}
\usepackage{graphicx}
\usepackage{dcolumn}
\usepackage{bm}

\newcommand{\sgn}{\,\mbox{\rm sgn}\,}

\newcommand{\rmd}{\mathrm{d}}

\newcommand{\llangle}{\langle\kern -.23em \langle}
\newcommand{\rrangle}{\rangle\kern -.23em \rangle}

\renewcommand{\vec}[1]{\boldsymbol{#1}}

\title{
  Linear Complexity Lossy Compressor for Binary Redundant Memoryless Sources
}
\author{
  \textsc{Kazushi MIMURA} 
  \thanks{E-mail address: \tt mimura@hiroshima-cu.ac.jp}
}
\inst{
  Graduate School of Information Sciences, Hiroshima City University,\\
  3--4--1 Ohtsuka-Higashi, Asaminami-ku, Hiroshima 731--3194, Japan
}
\abst{
  A lossy compression algorithm for binary redundant memoryless sources is presented. 
  The proposed scheme is based on sparse graph codes. 
  By introducing a nonlinear function, redundant memoryless sequences can be compressed. 
  We propose a linear complexity compressor based on the extended belief propagation, 
  into which an inertia term is heuristically introduced, 
  and show that it has near-optimal performance for moderate block lengths. 
}
\kword{
  nonlinear function, 
  LDGM codes, 
  lossy compression, 
  replica method
}

\begin{document}
\maketitle

%\section{Introduction}
%~~~~~~~~~~~~~~~~~~~~~~~~~~~~~~~~~~~~~~~~~~~~~~~~~~~~~~~~~~~~~~~~~~~~~
\par
Channel coding can be considered as a dual problem of lossy source coding \cite{Cover2006, Csiszar1981}. 
Recent research on error correcting codes and lossy source coding has shown 
that the statistical mechanical approach can be used to explain such problems \cite{Nishimori2001}. 
Lossy compression for memoryless sources has been widely investigated 
since Matsunaga and Yamamoto showed that it is possible to attain the rate-distortion bound asymptotically 
using low-density parity-check (LDPC) codes \cite{Matsunaga2003}．
The upper \cite{Dimakis2007} and lower \cite{Martinian2006, Wainwright2007} bounds 
on its rate-distortion performance of low-density generator-matrix (LDGM) codes for lossy compression  with a given check degree are evaluated. 
Some other lossy compression schemes that have asymptotic optimality have been proposed so far 
\cite{Murayama2003, Hosaka2002, Hosaka2006, Ciliberti2005, Ciliberti2006, Mimura2006, Gupta2009, Miyake2008a, Muramatsu2008, Cousseau2008a, Cousseau2008b, Mimura2009}．
\par
Efficient compressors, on the other hand, are still in the stage of development. 
Some efficient encoding algorithms, which have near optimal performance, have been proposed, 
e.g., 
the nested binary linear codes \cite{Wadayama2003}, 
the inertia-term-introduced belief propagation \cite{Murayama2004,Hosaka2006}，
the survey-propagation-based message passing algorithm \cite{Wainwright2005}，
the bit-flipping-based algorithm \cite{Gupta2009}, 
the exhaustive search of small words into what an original message is divided \cite{Gupta2008}，
the linear-programming-based algorithm, \cite{Miyake2008b} and 
the Markov-Chain Monte-Carlo(MCMC)-based algorithm \cite{Jalali2008}. 
\par
For redundant memoryless sources, some low complexity compressors, e.g., 
the near-linear complexity compressor based on the exhaustive search of small words \cite{Gupta2008}, 
the quadratic complexity compressor based on the bit-flipping-based algorithm \cite{Gupta2009}, 
and the MCMC-based compressor \cite{Jalali2008} whose complexity is independent of the sequence length, 
have been proposed so far. 
One of other approaches to obtain near-linear complexity compressors for redundant sources is to apply the inertia-term-introduced belief propagation. 
Hosaka and Kabashima have proposed an algorithm for redundant sources, whose complexity is $O(N^2)$ \cite{Hosaka2006}. 
In this study, we propose a linear complexity lossy compression algorithm based on an inertia-term-introduced belief propagation 
by using a nonlinear function and a sparse matrix such as low-density generator matrix (LDGM) codes for binary redundant memoryless sources. 
This proposed algorithm can be regarded as the perceptron-based one \cite{Hosaka2006} whose edges are extremely deleted to have a finite connectivity 
and has asymptotic optimality under some constraints. 
We also show that it has near optimal performance for moderate block lengths.

%\section{Background}
%~~~~~~~~~~~~~~~~~~~~~~~~~~~~~~~~~~~~~~~~~~~~~~~~~~~~~~~~~~~~~~~~~~~~~
\par
Let us first provide the concepts of the rate-distortion theory \cite{Cover2006} and some notations. 
Let $x$ be a discrete random variable with an alphabet $\mathcal{X}$. 
A source alphabet, a codeword alphabet, and a reproduced alphabet are $\mathcal{X}$, $\mathcal{S}$ and $\hat{\mathcal{X}}$, respectively. 
The compressor $\mathcal{F}$ encodes the $M$ bit source sequence $\vec{x}={}^t(x_1,\cdots,x_M) \in \mathcal{X}^M$ 
into the $N(<M)$ bit codeword $\vec{\xi}={}^t(\xi_1,\cdots,\xi_N)=\mathcal{F}(\vec{x})\in\mathcal{S}^N$. 
The decompressor $\mathcal{G}$ generates the $M$ bit reproduced sequence $\hat{\vec{x}}={}^t(\hat{x}_1,\cdots,\hat{x}_M)={\cal G}(\vec{\xi}) \in \hat{\cal X}^M$ from the codeword $\vec{\xi}$. 
The code rate then becomes $R=N/M(<1)$. 
\par
A distortion measure is a map $d : \mathcal{X} \times \hat{\mathcal{X}} \to [0,\infty)$. 
A distortion between the sequences $\vec{x} = {}^t (x_1, \cdots, x_M) \in \mathcal{X}^M$ 
and $\hat{\vec{x} } = {}^t (\hat{x}_1, \cdots, \hat{x}_M) \in \hat{\mathcal{X}}^M$ is measured 
by the averaged single-letter distortion as $d(\vec{x},\hat{\vec{x}}) = \frac 1M \sum_{k=1}^M d(x_k,\hat{x}_k)$. 
A rate distortion pair $(R,D)$ is said to be achievable 
if there exists a sequence of rate distortion codes $({\cal F},{\cal G})$ with $\mathbb{E}_{\vec{x}} [d(\vec{x},\hat{\vec{x}})] \le D$ in the limit $M\to\infty$. 
The rate distortion function $R(D)$ is the infimum of the rate $R$ such that $(R,D)$ is in the rate distortion region of the source for a given distortion $D$. 
\par
We hereafter consider the binary alphabets $\mathcal{X}=\mathcal{S}=\hat{\mathcal{X}}=\{ -1,1 \}$ 
and a redundant binary memoryless source whose distribution is given by $\mu(X=1)=1-p, \mu(X=-1)=p$. 
The parameter $p$ is a source bias. 
We use the Hamming distortion 
\begin{equation}
  d(x,\hat{x}) = \biggl\{
  \begin{array}{ll}
    0, & \; {\rm if} \; x = \hat{x}, \\
    1, & \; {\rm if} \; x \ne \hat{x},
  \end{array}
\end{equation}
as a distortion measure. 
%We use the Hamming distortion $d(x,\hat{x})$, which takes 0 if $x = \hat{x}$ , 1 otherwise, as a distortion measure. 
The rate-distortion function of a Bernoulli($1/2$) i.i.d. source then becomes $R(D)=h_2(p)-h_2(D)$, 
where $h_2$ denotes the binary entropy function which is defined by $h_2(x) = -x\log_2(x) - (1-x) \log_2(1-x)$.

%\section{Lossy compression scheme}
%~~~~~~~~~~~~~~~~~~~~~~~~~~~~~~~~~~~~~~~~~~~~~~~~~~~~~~~~~~~~~~~~~~~~~
\par
Using a parameter $\vec{w}=(w_1,w_2)$ $\in (\mathbb{N}\backslash \{0\})^2$, 
we first introduce the nonlinear function $g : \mathbb{N} \to \{-1,1\}$ defined as 
\begin{equation}
  g_{\vec{w}} (z) = g_{(w_1,w_2)} (z) = \left\{
  \begin{array}{rl}
     1 & \; \mathrm{if} \; w_1 < |z| < w_2 \\
    -1 & \; \mathrm{otherwise} 
  \end{array}
  \right. , 
\end{equation}
where the operator ${}^tA$ denotes the transpose of $A$. 
For a vector, this function acts componentwise. 
We here consider the following decompressor: 
\begin{equation}
  \mathcal{G}(\vec{\xi}) = g_{\vec{w}} ( \mathcal{A} \vec{\xi} ), 
\end{equation}
where $\mathcal{A}=(a_{\mu i}) \in \{-1, 0, 1\}^{M \times N}$ denotes a sparse regular matrix whose row weight, 
i.e., the number of nonzero elements, is $C$. 
Each nonzero element of the sparse matrix $\mathcal{A}$ takes $\pm 1$ with equiprobability. 
The function $g_{\vec{w}} (z)$ is adjusted 
so that the bias of the reproduced sequence is close to that of the original message as much as possible. 
More complex functions can be considered as the function $g_{\vec{w}} (z)$. 
As will be discussed later, the complexity of our proposed algorithm is $O(N)$ but it is proportional to $2^C$. 
So a small $C$ is preferable; therefore, we here consider a simple nonlinear function that can easily be adjusted. 
\par
The compressor is then defined by 
\begin{equation}
  \mathcal{F}(\vec{x}) = \mathop{\rm argmin}_{\vec{s} \in \{-1,1\}^N} d(\vec{x},\mathcal{G}(\vec{s})). 
  \label{eq:def_F}
\end{equation}
\par
This compressor is identical to $\mathcal{F}(\vec{x}) = \mathop{\rm argmax}_{\vec{s} \in \{-1,1\}^N}$ $p(\vec{s};\vec{x})$, 
which is the maximization of the following distribution 
\begin{equation}
p(\vec{s};\vec{x}) 
  \triangleq \frac 1{Z(\beta)} e^{-\beta M d(\vec{s};\vec{x})}
  = \frac 1{Z(\beta)} \prod_{k=1}^M e^{-\beta G_k(\vec{s};x_k)}, 
\end{equation}
where 
$G_k(\vec{s};x_k) = \frac 12 (1-x_k \hat{x}_k)$, 
$\hat{x}_k = g_{\vec{w}} ( \sum_{i \in \mathcal{L}(k)} a_{ki} s_i )$ 
and $\mathcal{L}(k) = \{i|a_{ki} \ne 0\}$ with the parameter $\beta >0$. 
Here, $Z(\beta)$ denotes an normalization constant of $p(\vec{s};\vec{x})$, 
which is defined by $Z(\beta) = \sum_{\vec{s}} e^{-\beta M d(\vec{x},\mathcal{G}(\vec{s}))}$. 
The function $G_k(\vec{s};x_k)$ represents a distortion with respect to the $k$ th bit.

%\section{Analysis in the large row weight limit}
%~~~~~~~~~~~~~~~~~~~~~~~~~~~~~~~~~~~~~~~~~~~~~~~~~~~~~~~~~~~~~~~~~~~~~
\par
We here consider a large row weight limit, which is a case where $C = N$ holds, to allows us to infer compression performance. 
In this limit our scheme can be regarded as the perceptron-based code \cite{Hosaka2002, Hosaka2006, Cousseau2008a, Cousseau2008b, Mimura2011}. 
When $w_2>N$, these are equivalent to each other. 
To make the parameters $w_1$ and $w_2$ be of order unity, 
we introduce a constant into the decompressor as $\mathcal{G}(\vec{\xi}) = \vec{g}_{\vec{w}} ( N^{-1/2} \mathcal{A} \vec{\xi} )$. 
The achievable distortion $D$ is evaluated as $D = \lim_{\beta \to \infty} \partial [\beta f(\beta)]/\partial \beta$ 
via the free energy density $f(\beta) = (-\beta M)^{-1} \mathbb{E}_{\mathcal{A},\vec{x}}[ \ln Z(\beta)]$, where $\mathbb{E}$ denotes an expectation operator. 
\par
Applying the so called replica method, the free energy density can be evaluated as 
$f(\beta) = - \beta^{-1} ( p \ln \{ e^{-\beta} + (1-e^{-\beta}) K_{\vec{w}} \} + (1-p) \ln \{ e^{-\beta} + (1-e^{-\beta}) (1-K_{\vec{w}}) \}  + R \ln 2)$ 
within the replica symmetric treatment, where 
$K_{\vec{w}} = \int_{\{z \in \mathbb{R} |g_{\vec{w}}(z)=-1\}} {(2\pi)}^{-1/2} e^{-z^2/2} \rmd z$. 
The parameter $K_{\vec{w}}$ is identical to the expectation value $\mathbb{E}_z[g_{\vec{w}}(z)=-1]$ with a random variable $z$ which obeys the standard normal distribution $\mathcal{N}(0,1)$, 
which originates from the distribution of each element of $N^{-1/2} \mathcal{A} \vec{\xi}$. 
It can be considered that the compression performance is given using the distribution of $N^{-1/2} \mathcal{A} \vec{\xi}$ and $g_{\vec{w}}$ in this scheme. 
\par
The entropy density of $p(\vec{s};\vec{x})$ is then obtained as $s(\beta) = \beta ( \partial [\beta f(\beta)]/\partial \beta - f )$. 
The entropy density $s(\beta)$ must be non-negative owing to the definition of $p(\vec{s};\vec{x})$; 
however it takes negative values in the large $\beta$ region. 
We therefore evaluate the achievable distortion $D$ at $\beta_c$ which gives a zero entropy density ($s(\beta_c)=0$), 
so this analysis is equivalent to the Krauth-M\'ezard approach which is a kind of one-step replica symmetric breaking (RSB) treatment \cite{Krauth1989}. 
\par
Using the zero-entropy-density criterion, 
minimizing the achievable distortion $D=\lim_{\beta \to \beta_c} \partial [\beta f(\beta)]/\partial \beta$ with respect to $\vec{w}$, 
one can obtain $R(D)=h_2(p)-h_2(D)$, which is identical to the rate-distortion function. 
From these two conditions, i.e., the zero entropy density and the minimization of the achievable distortion, the following relationships are obtained: 
\begin{eqnarray}
  & & e^{-\beta} = \frac D{1-D}, \label{eq:condition1} \\
  & & K_{\vec{w}} = \frac{p-D}{1-2D}. \label{eq:condition2} 
\end{eqnarray}

%\section{Linear complexity message passing compressor}
%~~~~~~~~~~~~~~~~~~~~~~~~~~~~~~~~~~~~~~~~~~~~~~~~~~~~~~~~~~~~~~~~~~~~~
\par

%\subsection{Basis}
%~~~~~~~~~~~~~~~~~~~~~~~~~~~~~~~~~~~~~~~~~~~~~~~~~~~~~~~~~~~~~~~~~~~~~
\par
The definition of the compressor means that it has exponential complexity. 
We then utilize a suboptimal algorithm based on message passing to construct the compressor \cite{Murayama2004}. 
Instead of the maximization of $p(\vec{s};\vec{x})$ we use a symbol MAP encoding scheme, which is maximization of marginal distribution, 
\begin{equation}
  \xi_i = \mathop{\rm argmax}_{s_i \in \{-1,1\} } \sum_{\vec{s} \backslash s_i \in \{-1,1\}^{N-1} } p(\vec{s};\vec{x}). 
\end{equation}
To evaluate the marginal distribution we apply the belief propagation. 
Since $\mathcal{G}({-\vec{s}})=\mathcal{G}({\vec{s}})$ holds for any $\vec{s}$, the expectation value of $s_i$ becomes zero. 
To avoid this uncertainty we heuristically introduce an inertia term as a prior, which gives the following {\it inertia-term-introduced belief propagation} \cite{Murayama2004}: 
\begin{eqnarray}
  & & \hspace*{-5mm} \hat{\rho}_{ki}^t(s_i) = \sum_{s_{i' \in \mathcal{L}(k) \backslash i}} e^{-\beta G_k(\vec{s};x_k)} \prod_{i' \in \mathcal{L}(k) \backslash i} \rho_{i'k}^t(s_{i'}), \\
  & & \hspace*{-5mm} \rho_{ik}^{t+1}(s_i) = \alpha_{ik} r_i^t(s_i) \prod_{k' \in \mathcal{M}(i) \backslash k} \hat{\rho}_{k'i}^t(s_i). 
\end{eqnarray}
A pseudo marginal can be evaluated as 
\begin{equation}
  q_i^{t+1}(s_i) = \alpha_i r_i^t(s_i) \prod_{k \in \mathcal{M}(i)} \hat{\rho}_{ki}^t(s_i), 
\end{equation}
where $\alpha_{ik}$ and $\alpha_i$ denote normalization constants 
and $\mathcal{M}(i)=\{k|a_{ki} \ne 0 \; \forall \mu \}$. 
Here, the function $r_i^t(s_i)$ is an introduced prior as the inertia term and the superscript $t$ represents an iteration step. 
We here consider that $s_i$ is binary, so we can safely put 
$\rho_{ik}^t(s_i) = \frac12(1 + m_{ik}(t) s_i)$, 
$\hat{\rho}_{ki}^t(s_i) = \frac12(1 + \hat{m}_{ik}(t) s_i)$, 
$q_i^t(s_i) = \frac12(1 + m_i(t) s_i)$. 
We here define a prior as $r_i^t(s_i) = e^{s_i \tanh^{-1} [\gamma m_i(t)]}$, 
where the parameter $\gamma \; (0\le\gamma<1)$ denotes the amplitude of the inertia term, which is heuristically chosen. 
When $\gamma=0$ ($r_i^t(s_i)=1$), the inertia-term-introduced belief propagation recovers the conventional belief propagation. 
It should be noted that the performance does not strongly depend on the detailed shape of the function, if it is an increasing function. 
It has not yet been investigated how the inertia term works in detail so far; 
however it is known that the inertia term chooses a single peak in the calculation of the pseudo marginal.

%\subsection{Encoding algorithm}
%~~~~~~~~~~~~~~~~~~~~~~~~~~~~~~~~~~~~~~~~~~~~~~~~~~~~~~~~~~~~~~~~~~~~~
\par
We calculate the equations of the belief propagation, which gives 
\begin{eqnarray}
  \hat{m}_{ki}(t) 
  &=& \frac{\displaystyle a_{ki} x_k \biggl(\tanh \frac{\beta}2 \biggr) V_{ki}(t)} {\displaystyle 1 + x_k \biggl(\tanh \frac{\beta}2 \biggr) U_{ki}(t)}, \label{eq:algo1} \\
  m_{ik}(t+1)
  &=& \tanh \Biggl( \sum_{k' \in \mathcal{M}(i) \backslash k} \tanh^{-1} \hat{m}_{k'i}(t) \nonumber \\
  & & \qquad\qquad + \tanh^{-1} \gamma m_i(t) \Biggr), \\ 
  m_i(t+1)
  &=& \tanh \Biggl( \sum_{k' \in \mathcal{M}(i)} \tanh^{-1} \hat{m}_{ki}(t) \nonumber \\
  & & \qquad\qquad + \tanh^{-1} \gamma m_i(t) \Biggr), 
\end{eqnarray}
where 
\begin{eqnarray}
  U_{ki}(t) 
  &=& \sum_{s_{i' \in \mathcal{L}(k) \backslash i}} 
      u_{\vec{w}} \Biggl( \sum_{i' \in \mathcal{L}(k) \backslash i} a_{ki'} s_{i'} \Biggr) \nonumber \\
  & & \times \prod_{i' \in \mathcal{L}(k) \backslash i} 
      \frac{1+m_{i'k}(t)s_{i'}}2, \\ 
  V_{ki}(t) 
  &=& \sum_{s_{i' \in \mathcal{L}(k) \backslash i}} 
      v_{\vec{w}} \Biggl( \sum_{i' \in \mathcal{L}(k) \backslash i} a_{ki'} s_{i'} \Biggr) \nonumber \\
  & & \times \prod_{i' \in \mathcal{L}(k) \backslash i} \!\!\!\! 
      \frac{1+m_{i'k}(t)s_{i'}}2, \\
  u_{\vec{w}}(x) 
  &=& \mathbb{I}( -w_2 < x < -w_1 ) + \mathbb{I}( w_1 < x < w_2) \nonumber \\
  & & - \mathbb{I}(x<-w_2) - \mathbb{I}(w_2<x) \nonumber \\
  & & - \mathbb{I}(-w_1<x<w_1), \\
  v_{\vec{w}}(x) 
  &=& \mathbb{I}(x=-w_2) - \mathbb{I}(x=-w_1) \nonumber \\
  & & + \mathbb{I}(x=w_1) - \mathbb{I}(x=w_2), 
\end{eqnarray}
and $\mathbb{I}(\mathcal{P})$ denotes an indicator function that takes 1 if the proposition $\mathcal{P}$ is true, and 0 otherwise. 
After $t_m$ iterations, the $i$ th bit of the codeword can be obtained as $\xi_i = \sgn[m_i(t_m)]$ using the mean of the pseudo marginal $m_i(t_m)$. 
To derive these iterative equations, we use the identity 
$\hat{x}_k = u_{\vec{w}} ( \sum_{i' \in \mathcal{L}(k) \backslash i} a_{ki'} s_{i'} ) 
+ a_{ki} s_i \; v_{\vec{w}} ( \sum_{i' \in \mathcal{L}(k) \backslash i} a_{ki'} s_{i'} )$, 
which holds for any $i \in \mathcal{L}(k)$. 
\par
The computational cost of the terms $U_{ki}(t)$ and $V_{ki}(t)$ is $O(2^C)$, which depends only on the row weights $C$, 
namely, it is $O(1)$ with respect to $N$. 
The complexity of this algorithm is therefore $O(N)$ when the number of iterations $t_m$ is fixed.

%\subsection{Parameter settings}
%~~~~~~~~~~~~~~~~~~~~~~~~~~~~~~~~~~~~~~~~~~~~~~~~~~~~~~~~~~~~~~~~~~~~~
\par
Utilizing eqs. (\ref{eq:condition1}) and (\ref{eq:condition2}) which are obtained in the large-row-weight-limit analysis, 
we can approximately set all parameters $C$, $w_1$, $w2$, and $\beta$ of our scheme with finite row weights except $\gamma$. 
\par
We first consider a setting of the parameter $\beta$. 
Using eq. (\ref{eq:condition1}) and the rate-distortion function, 
we set $\beta$ as $\beta=\beta_c(p,R)$ for the given the source bias $p$ and the code rate $R$, where 
$\beta_c (p,R) = \ln ( {[h_2^{-1}(h_2(p)-R)]}^{-1} -1 )$. 
Here, $h_2^{-1}$ denotes the inverse function of the binary entropy function. 
\par
We next consider a setting of the parameters $C$, $w_1$, and $w2$. 
Each element of the vector $\mathcal{A} \vec{s}$ is the summation of $C$ Bernoulli random variables $1-2Ber(0.5)$, where $\vec{s} \in \{-1,1\}^N$ denotes a candidate of a codeword. 
This is a similar situation to the row weight limit. 
To keep the row weight finite, we restrict the row weight as $C \le C_{max}$. 
Using eq. (\ref{eq:condition2}) and the rate-distortion function, we set $(C,w_1,w_2)$ as 
$(C,w_1,w_2) = \mathop{\rm argmin}_{(C',w_1',w_2') \in D(C_{max})} |\hat{K}(C',w_1',w_2') - K(p,R)|$, 
where 
$\hat{K}(C,w_1,w_2)$ $= \sum_{n\in\{0,\cdots,C\}: g_{\vec{w}}(C-2n)=-1} 2^{-C} (_{\, n}^C)$, 
$K(p,R) = [p-h_2^{-1}(h_2(p)-R)]/[1-2h_2^{-1}(h_2(p)-R)]$, 
and $D(C_{max})=\{ (C,w_1,w_2)| 2 \le C \le C_{max}, 0 < w_1 \le C-1, w_1<w_2 \le C+1 \}$ for the given $p$ and $R$. 
Note that the parameter that gives second smallest value might provide better performance. 
\par
Lastly, $\gamma$ is determined by trial and error. 
In this study, we choose $\gamma$ only within $\{0.2, 0.3, 0.4, 0.5\}$.

%\section{Experiments}
%~~~~~~~~~~~~~~~~~~~~~~~~~~~~~~~~~~~~~~~~~~~~~~~~~~~~~~~~~~~~~~~~~~~~~
\par
\begin{figure}[t]%[htbp]
  \begin{center}
    \includegraphics[width=.8\linewidth,keepaspectratio]{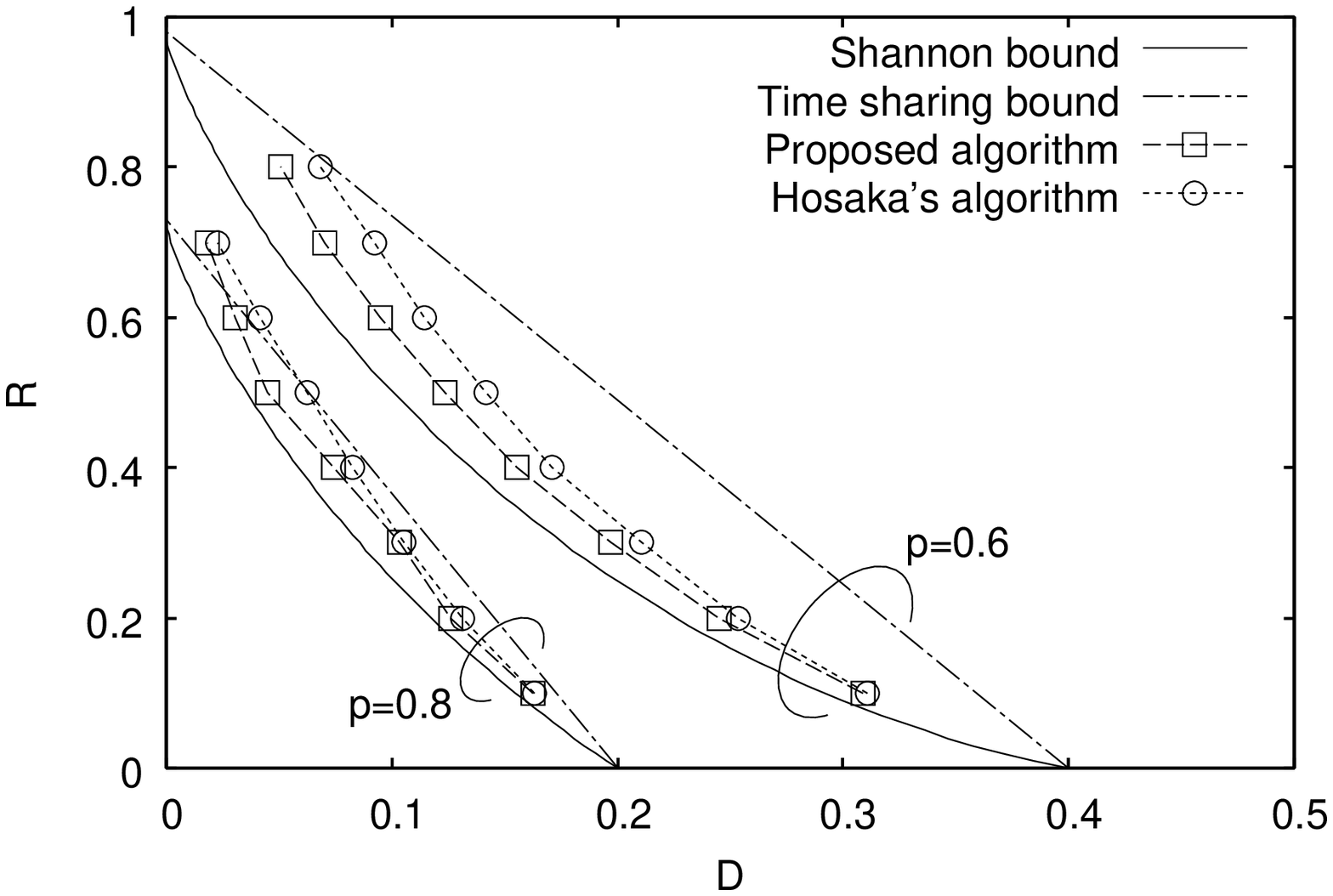} \\[3mm]
    \includegraphics[width=.8\linewidth,keepaspectratio]{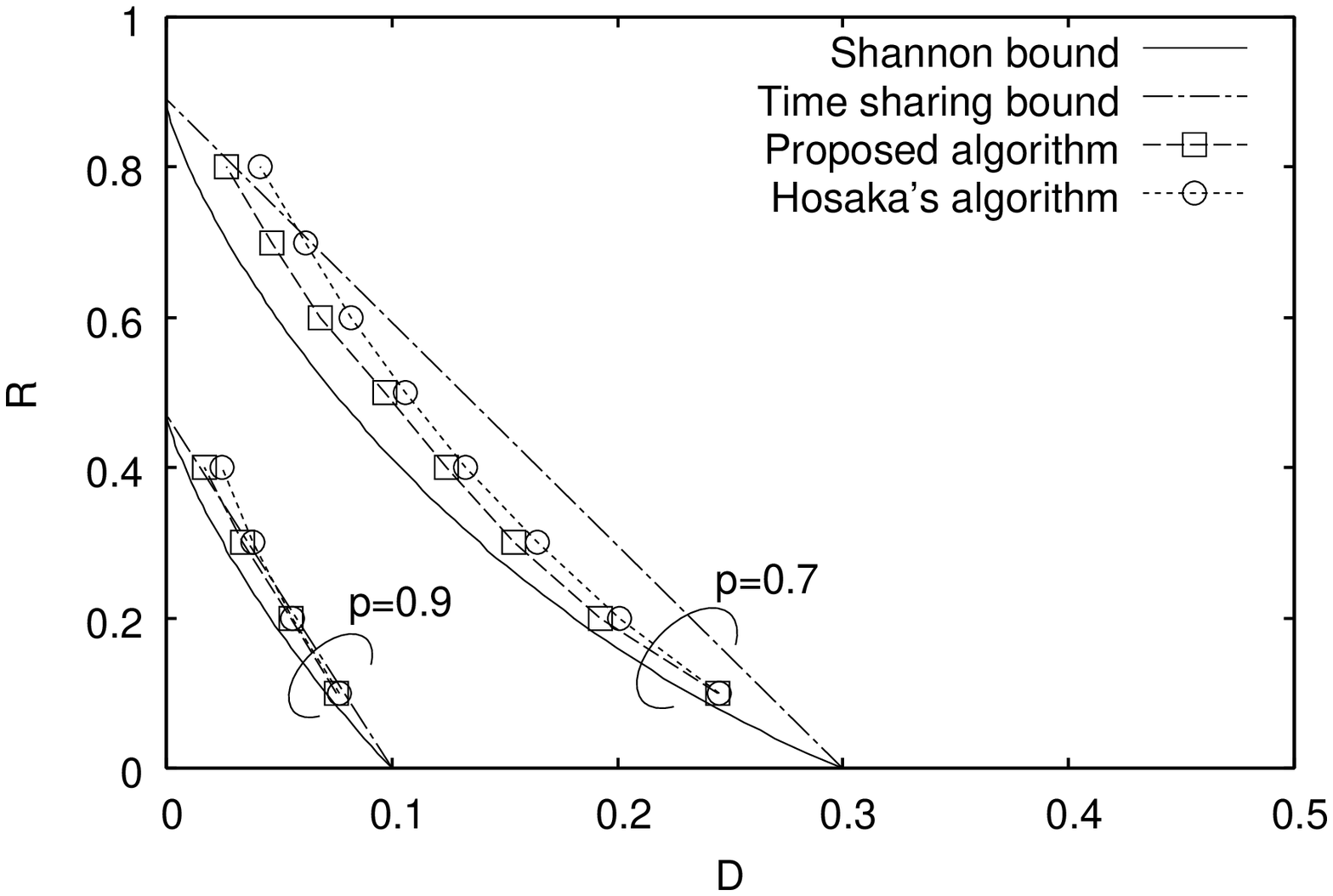} \\
    \caption{
      Empirical compression performance against the code rate $R$ for typical source bias $p$. 
      The proposed algorithm (squares) and Hosaka's algorithm (circles) are shown. 
      The length of the orginal sequences is $N=420$, and all the measurements are averaged over 10 runs. 
      The parameter $\gamma$ is chosen within $\{0.2, 0.3, 0.4, 0.5\}$. 
      The row weight $C$ is chosen within $C \le 8$ ($C_{max}=8$). 
      Top: $p \in \{0.6, 0.8\}$. 
      Bottom: $p \in \{0.7, 0.9\}$. 
    }
    \label{fig:performance}
  \end{center}
\end{figure}
\par
The empirical compression performance is shown in Fig. \ref{fig:performance}. 
In this figure, the distortion averaged over $10$ runs is plotted as a function of the code rate $R$ for the source bias $p \in \{ 0.6, 0.7, 0.8, 0.9 \}$. 
The length of codewords is fexed at $N=420$, and the length of original sequence is adjusted. 
We here choose $C_{max}=8$ and $p>0.5$. 
In this figure, the time sharing bound is also shown. 
The time sharing bound is given by $R(D) = ( 1- \frac Dp) h_2(p)$, 
which denotes the compression performance 
achieved by the time sharing scheme of a lossless coding ($R(0)=h_2(p)$)
and a trivial encoding that always outputs an all-one vector for any input ($R(p)=0$). 
It can be confirmed that the proposed linear complexity compressor (with $O(N)$ complexity) 
has slightly better performance than Hosaka's algorithm (with $O(N^2)$ complexity). 
\par
We observed a not-so-good performance for the small-$p$ region that the source bias is less than about $0.2$. 
In this region $\mathrm{min} |\hat{K}(C',w_1',w_2')-K(p,R)|$ is not much smaller than that of the large-$p$ region. 
When we compress the original sequence $\vec{x}$ of which bias is $p<0.5$, 
we can first flip it as $-\vec{x}$ and then compress. 
The information for determining whether the sequence flips requires one bit. 
To reduce $\mathrm{min} |\hat{K}(C',w_1',w_2')-K(p,R)|$, it might be helpful to introduce more complex nonlinear functions.

%\section{Conclusion}
%~~~~~~~~~~~~~~~~~~~~~~~~~~~~~~~~~~~~~~~~~~~~~~~~~~~~~~~~~~~~~~~~~~~~~
\par
In this study, we have proposed a scheme using a nonlinear function and a sparse matrix, 
and as well as a linear complexity message passing compressor based on the inertia-term-introduced belief propagation. 
The proposed method can treat redundant memoryless sources and has near-optimal compression performance for moderate block lengths. 
The adjustment of the column weight distribution of the sparse matrix might enable us to improve the compression performance. 
The analysis of this scheme with finite row weights is one of our future studies.

%\section*{Acknowledgment}
%~~~~~~~~~~~~~~~~~~~~~~~~~~~~~~~~~~~~~~~~~~~~~~~~~~~~~~~~~~~~~~~~~~~~~
The author would like to thank Tadaaki Hosaka, Tatsuto Murayama, and Jun Muramatsu for their valuable comments and 
Ankit Gupta, Sergio Verd\'u, and Tsachy Weissman for providing him valuable preprints. 
This work was partially supported by 
a Grant-in-Aid for Scientific Research (C) No. 22500136 
from the Ministry of Education, Culture, Sports, Science and Technology (MEXT) of Japan.


\begin{thebibliography}{99}
%~~~~~~~~~~~~~~~~~~~~~~~~~~~~~~~~~~~~~~~~~~~~~~~~~~~~~~~~~~~~~~~~~~~~~


% origin 

\bibitem{Cover2006} 
T. M. Cover and J. A. Thomas: 
{\it Elements of Information Theory, 2nd ed.}, 
John Wiley and Sons Inc., 2006.

\bibitem{Csiszar1981} 
I. Csisz\'ar and J. K\"orner: 
{\it Information Theory: Coding Theorems for Discrete Memoryless Systems}, 
Academic Press, 1981. 

\bibitem{Nishimori2001}
H. Nishimori: 
Statistical Physics of Spin Glasses and Information Processing -- An Introduction, 
Oxford University Press, 2001. 

\bibitem{Matsunaga2003}
Y. Matsunaga and H. Yamamoto: 
%"A Coding Theorem for Lossy Data Compression by LDPC codes," 
{\it IEEE Trans. Inform. Theory}, vol. 49, 2225, 2003. 


% analytical evaluation of the rate-distortion performance

\bibitem{Dimakis2007}
A. G. Dimakis, M. J. Wainwright, and K. Ramchandran: 
%"Lower bounds on the rate-distortion function of LDGM codes," 
{\it Info. Theory Workshop}, 650, 2005. 

\bibitem{Martinian2006}
E. Martinian and M. J. Wainwright: 
%"Analysis of LDGM and compound codes for lossy compression and binning," 
{\it Workshop on Information Theory and its Applications}, 2006. 

\bibitem{Wainwright2007} 
M. J. Wainwright: 
%"Sparse Graph Codes for Side Information and Binning," 
{\it IEEE Signal Processing Magazine}, 47, Sept. 2007. 

% proposal of a scheme

\bibitem{Murayama2003}
T. Murayama and M. Okada: 
%"One step RSB scheme for the rate-distortion function,"  
{\it J. Phys. A: Math. Gen.}, vol. 36, 11123, 2003. 

\bibitem{Hosaka2002}
T. Hosaka, Y. Kabashima, and H. Nishimori: 
{\it Phys. Rev. E}, 66, 066126, 2002. 

\bibitem{Hosaka2006}
T. Hosaka and Y. Kabashima: 
%"Statistical mechanical approach to lossy data compression: Theory and Practice," 
{\it Physica A}, 365, 113, 2006. 

\bibitem{Ciliberti2005}
S. Ciliberti and M. M\'ezard: 
%"The Theoretical capacity of the Parity Source Coder," 
{\it J Stat. Mech.}, vol. 3, 58, 2006. 

\bibitem{Ciliberti2006}
S. Ciliberti, M. M\'ezard, and R. Zecchina: 
%"Message-Passing Algorithms for Non-Linear Nodes and Data Compression," 
{\it Complex Syst. Methods}, vol. 3, 58, 2006. 

\bibitem{Mimura2006}
K. Mimura and M. Okada: 
%"Statistical mechanics of lossy compression using multilayer perceptrons," 
{\it Phys. Rev. E}, 74, 026108, 2006. 

%\bibitem{Gupta2007}
%A. Guputa and S. Verd\'u, 
%%"Nonlinear Sparse-Graph Codes for Lossy Compression of Discrete Nonredundant Sources,"
%{\it Info. Theory Workshop 2005}, 1493, 2005. 

\bibitem{Gupta2009}
A. Gupta and S. Verdu: 
%``Nonlinear Sparse-Graph Codes for Lossy Compression,'' 
{\it IEEE Trans. Inform. Theory}, vol. 55, 1961, 2009. 

\bibitem{Miyake2008a}
S. Miyake and J. Muramatsu: 
%"A construction of Lossy source code using LDPC matrices," 
{\it IEICE Trans. Fundamentals}, E91-A, 1488, 2008. 

\bibitem{Muramatsu2008}
J. Muramatsu: 
{\it Proc. 2008 IEEE Int'l. Sympo. Info. Theory (ISIT2008)}, 424, 2008. 

\bibitem{Cousseau2008a}
F. Cousseau, K. Mimura, T. Omori, and M. Okada: 
%"Statistical mechanics of lossy compression for nonmonotonic multilayer perceptrons," 
{\it Phys. Rev. E}, 78, 021124, 2008.

\bibitem{Cousseau2008b}
F. Cousseau, K. Mimura, and M. Okada: 
{\it Proc. 2008 IEEE Int'l. Sympo. Info. Theory (ISIT2008)}, 509, 2008. 

\bibitem{Mimura2009}
K. Mimura: 
%"Typical performance of irregular low-density generator-matrix codes for lossy compression," 
{\it J. Phys. A: Math. Theor.}, 42, 135002, 2009.




% compressor

\bibitem{Wadayama2003}
T. Wadayama, 
%"A lossy compression algorithm using an LDPC code for bianry iid sources", 
{\it Proc. of 3rd Int'l Sympo. Turbo Codes and Relatd Topics}, 231, 2003. 

\bibitem{Murayama2004}
T. Murayama, 
%"Thouless-Anderson-Palmer approach for lossy compression," 
{\it Phys. Rev. E}, vol. 69, 035105(R), 2004. 

\bibitem{Wainwright2005} 
M. J. Wainwright and E. Maneva, 
%"Lossy source encoding via messaga-passing and decimation over generalized codewords of LDGM codes," 
{\it Proc. 2008 IEEE Int'l. Sympo. Info. Theory (ISIT2008)}, 1493, 2005. 

\bibitem{Gupta2008}
A. Gupta, S. Verdu, and T. Weissman, 
%``Rate-distortion in near-linear time,'' 
{\it Proc. 2008 IEEE Int'l. Sympo. Info. Theory (ISIT2008)}, 847, 2008. 

\bibitem{Miyake2008b}
S. Miyake, J. Honda, and H. Yamamoto, 
%"Application of LCPC to Lossy Source Coding," 
{\it Proc. 2008 IEEE Int'l. Sympo. Info. Theory and its Applications (ISITA2008)}, 589, 2008. 

\bibitem{Jalali2008}
S. Jalali and T. Weissman, 
%``Rate-distortion via Markov chain Monte Carlo,''
{\it Proc. 2008 IEEE Int'l. Sympo. Info. Theory (ISIT2008)}, 852, 2008. 

\bibitem{Mimura2011}
K. Mimura, F. Cousseau, and M. Okada, 
%``Belief propagation for error correting codes and lossy compression using multilayer perceptrons,'' 
{\it J. Phys. Soc. Jpn.}, 80, 3, 034802, 2011. 

\bibitem{Krauth1989}
W. Krauth and M. M\'ezard, 
{\it J. Phys. France}, 50, 3057, 1989. 


\end{thebibliography}
\end{document}